\documentclass[letterpaper, 10 pt, conference]{ieeeconf}  % Comment this line out if you need a4paper

\IEEEoverridecommandlockouts                              % This command is only needed if 
\usepackage{cite}
\usepackage{amsmath,amssymb,amsfonts}
\usepackage{algorithmic}
\usepackage{graphicx}
\usepackage{textcomp}
\usepackage{subcaption}
\usepackage{lipsum}
\usepackage{mathtools}
\usepackage{cuted}
\usepackage{stfloats}
\usepackage{enumerate}
\usepackage[hidelinks]{hyperref}
\usepackage{amsmath,amssymb,amsfonts}
\def\Ac{{\mathcal A}}

\def\Dc{{\mathcal D}}

\def\Ec{{\mathcal E}}

\def\Gc{{\mathcal G}}

\def\Lc{{\mathcal L}}

\def\Mc{{\mathcal M}}
\def\Mbb{{\mathbb M}}

\def\Nc{{\mathcal N}}

\def\Pbb{{\mathbb P}}

\def\Rbb{{\mathbb R}}

\def\Vc{{\mathcal V}}

\def\th{{\theta}}

\def\Om{{\Omega}}

\def\0{{\bf 0}}

\def\underdotline#1{{\setbox0=\hbox{#1}\rlap{\raise-0.28em\hbox to\wd0{\rm\tiny\cleaders\hbox{.\kern-0.1ex}\hfill}}\box0}}
\def\underdashline#1{{\setbox0=\hbox{#1}\rlap{\raise-0.28em\hbox to\wd0{\rm\tiny\cleaders\hbox{-\kern-0.1ex}\hfill}}\box0}}
\def\understarline#1{{\setbox0=\hbox{#1}\rlap{\raise-0.28em\hbox to\wd0{\rm\tiny\cleaders\hbox{$\star$\kern-0.1ex}\hfill}}\box0}}

\newcommand{\bitem}{\begin{itemize}}
\newcommand{\eitem}{\end{itemize}}
\newcommand{\btabular}{\begin{tabular}}
\newcommand{\etabular}{\end{tabular}}
\newcommand{\bcenter}{\begin{center}}
\newcommand{\ecenter}{\end{center}}
\newcommand{\bea}{\begin{eqnarray}}
\newcommand{\eea}{\end{eqnarray}}
\newcommand{\bean}{\begin{eqnarray*}}
\newcommand{\eean}{\end{eqnarray*}}
\newcommand{\ba}{\left[ \begin{array}}
\newcommand{\ea}{\\ \end{array} \right]}
\newcommand{\bear}{\begin{array}}
\newcommand{\eear}{\\ \end{array}}

\usepackage{xcolor}

\newcommand{\noi}{\noindent}
\newcommand{\non}{\nonumber}

\newcommand{\bs}{\boldsymbol}

\font\myownfont=cmr17 scaled \magstep5
\def\psfancypar#1#2{\def\biginitial#1{{\myownfont#1}}%
  \def\makeinitial#1{\setbox8\hbox{\strut\vbox to 1.3ex
    {\hbox{\biginitial#1}\vskip -4pc plus 3.5pc minus 3.5pc}}}%
  \makeinitial#1%
  \ifdim\parindent>1.3\wd8\dimen8=\parindent
     \else\dimen8=1.3\wd8\fi
  \hangindent=\dimen8\hangafter=-2
  \noindent
  \strut\hskip-1\dimen8\box8{\sc#2}}%

\newcounter{subequation}
\def\beasub{\addtocounter{equation}{+1}
\setcounter{subequation}{\value{equation}}
\setcounter{equation}{0}
\renewcommand{\theequation}{\arabic{subequation}\alph{equation}}
\begin{eqnarray}}
\def\eeasub{\end{eqnarray}
\setcounter{equation}{\value{subequation}}
\renewcommand{\theequation}{\arabic{equation}}}

\newtheorem{Remark}{Remark}
\newtheorem{Lemma}{Lemma}

\newtheorem{Definition}{Definition}
\newtheorem{Theorem}{Theorem}

\title{\LARGE \bf
Collision-free Formation Control of Multiple Nano-quadrotors
}

\author{Anh Tung Nguyen$^{1}$, Ji-Won Lee$^{2}$, Thanh Binh Nguyen$^{3}$, and Sung Kyung Hong$^{1}$% <-this % stops a space
\thanks{*This work was supported by the MSIT (Ministry of Science and ICT), Korea, under the ITRC (Information Technology Research Center) support program (IITP-2021-2018-0-01424) supervised by the IITP (Institute for Information \& communications Technology Promotion). 
{\it (Corresponding author: Sung Kyung Hong.)}
}% <-this % stops a space
\thanks{$^{1}$Anh Tung Nguyen and Sung Kyung Hong are with 
		Faculty of Mechanical and Aerospace Engineering, Sejong University, Seoul 143--747(05006), Korea
        {\tt\small tung2610@sju.ac.kr; skhong@sejong.ac.kr}.}%
\thanks{$^{2}$Ji-Won Lee is with Human-Robot Interaction Research Center, Korea Institute of Robotics and Technology Convergence, Pohang 37553, Korea
	{\tt\small jiwon2@kiro.re.kr}}%
\thanks{$^{3}$Thanh Binh Nguyen is with Department of Control Engineering and Automation, Thuyloi University, 175 Tay Son, Dong Da, Hanoi, Vietnam
        {\tt\small ntbinh@tlu.edu.vn}}%
}

\begin{document}

\maketitle
\thispagestyle{empty}
\pagestyle{empty}

%%%%%%%%%%%%%%%%%%%%%%%%%%%%%%%%%%%%%%%%%%%%%%%%%%%%%%%%%%%%%%%%%%%%%%%%%%%%%%%%
\begin{abstract}
The utilisation of unmanned aerial vehicles has witnessed significant growth in real-world applications including surveillance tasks, military missions, and transportation deliveries.
This letter investigates practical problems of formation control for multiple nano-quadrotor systems.
To be more specific, the first aim of this work is to develop a theoretical framework for the time-varying formation flight of the multi-quadrotor system regarding anti-collisions.
In order to achieve this goal, the finite cut-off potential function is devoted to avoiding collisions among vehicles in the group as well as between vehicles and an obstacle.
The control algorithm navigates the group of nano-quadrotors to asymptotically reach an anticipated time-varying formation.
The second aim is to implement the proposed algorithm on Crazyflies nano-quadrotors, one of the most ubiquitous indoor experimentation platforms.
Several practical scenarios are conducted to tendentiously expose anti-collision abilities 
among group members as well as between vehicles and an obstacle.
The experimental outcomes validate the effectiveness of the proposed method in the formation tracking and the collision avoidance of multiple nano-quadrotors.
\end{abstract}
%%%%%%%%%%%%%%%%%%%%%%%%%%%%%%%%%%%%%%%%%%%%%%%%%%%%%%%%%%%%%%%%%%%%%%%%%%%%%%%%
\begin{keywords}
Multi-Robot Systems, Formation Control, Collision Avoidance, Obstacle Avoidance, Nano-quadrotors.
\end{keywords}

%%%%%%%%%%%%%%%%%%%%%%%%%%%%%%%%%%%%%%%%%%%%%%%%%%%%%%%%%%%%%%%%%%%%%%%%%%%%%%%%
\section{Introduction}
Over the past few decades, multi-robot systems have gained massive popularity in industrial societies due to their ability to describe large-scale interconnected systems in a variety of real-world applications such as transportation systems \cite{chen2010review,li2017distributed} and power systems \cite{zhao2012energy}.
In the multi-robot systems, the formation control problem aims at steering multiple robots in a network to achieve and maintain their predefined geometric patterns in their states, posing an immense challenge to the scientific community.
Since a colossal number of versatile robotic applications are developed, there have been a great deal of the works devoted to unmanned aerial vehicles (UAVs) \cite{li2016receding}, ground mobile robots \cite{dong2016time}, and especially to  
formation control of mobile robots \cite{bayezit2012distributed}.
%%

%%
%Especially, 
Quadrotors, one of the most ubiquitous classes of UAVs, have been intensively developed in many broad applications for assisting humans in difficult missions or hazard environments \cite{mogili2018review,nguyen2019quadcopter, nguyen2019active, nguyen2020dynamic}, e.g., in agriculture \cite{yallappa2017development}, industry \cite{benito2014design}, and military \cite{erdos2013experimental}.
Among numerous commercial products of quadrotors, this letter mainly focuses on studying nano-quadrotors (a miniature dimension of quadrotors).
The use of nano-quadrotors tolerates a convenience and simply installed experiments to verify control algorithms as well as conducting new research \cite{honig2018trajectory,luis2020online}.
In addition, actual flight tests on the nano-quadrotors also can be deployed in incommodious applications, and suitable for most laboratory setups.
After successfully validating control algorithms on the nano-quadrotors, researchers possibly scale up to other sizeable %sizeable có nghĩa là lớn a ơi
quadrotors in dealing with a colossal number of civilian applications.

Recently, collisions among robots have become a vital issue
when autonomously operating multi-robot systems, i.e., collisions among group members, and collisions between robots and obstacles.
Based on actual applications
%% cite deo gi nhieu the, 
%em cite xong e nhắc lại kĩ hơn ở dưới mà
\cite{wood2020collision,mansouri2020unified,liao2016distributed,dong2014time,ang2018high,nguyen2020distributed}, the desired spots of robots are generally marked with a view to guaranteeing a safe distance among group members.
Because of this arrangement, collisions among vehicles possibly occur when they move from their initial positions to the desired spots to result in an anticipated formation.
%% chỗ này k ổn à a?
%%Trajectories connecting such locations of vehicles highly probably cross each other, leading to unforeseen collision points.
%%
%%When at least two vehicles reach these points concurrently, they are likely to crash and be deprived of finalising their missions.
%%
Additionally, after completing the given formation, the group of robots may track a predefined trajectory.
This task is probably unsuccessful when obstacles appear in the trajectory and cause collisions with the robots.
With the purpose of dealing with the collision avoidance problem, path planning based on optimisation problems was presented in \cite{wood2020collision,lindqvist2020nonlinear,mansouri2020unified}.
Due to the computational cost of solving the optimisation problems, outcomes of such existing studies were limited by numerical results or operating a vehicle.
A coverage planning for ground mobile robots was introduced in \cite{kan2020online} to solve a problem of obstacle-cluttered environments in the two-dimensional space.
The potential field approach was developed to tackle a collision issue in \cite{boldrer2020socially,liao2016distributed,nguyen2020distributed,binh2020flocking}.
The authors in \cite{liao2016distributed,dong2014time} showed outdoor experimental results of formation control of multiple UAVs with consideration to inter-vehicle collisions.
Due to unclear collision points, the demonstrations of their algorithms would be vague. 

Motivated by the above observations, this letter is concerned with the formation control problem of multiple nano-quadrotor systems.
With consideration to the aforementioned collision issues, a collision-free formation control algorithm is proposed based on potential functions.
Accordingly,
our approach develops a new control Lyapunov function by which 
the multiple nano-quadrotor systems are stable
and the formation tracking errors exponentially converge.
In summary, our main contributions can be highlighted as follows.
\begin{enumerate}[i)]
	\item 
	This paper addresses a collision-free problem in a group of nano-quadrotors as well as between nano-quadrotors and an obstacle.
	A novel potential function is proposed to avoid local minima phenomenons.
	In addition, this function also guarantees the smoothness of the control input that increases practical capabilities.
	% of the group of aerial vehicles.
	%%
	\item 
	Based on the control Lyapunov function, a collision-free formation protocol is presented.
	As a result, the proposed protocol ensures that the group of nano-quadrotors asymptotically tracks the anticipated formation trajectory with no collisions.
	\item 
	It is worth nothing that the scope of our study mainly focuses on indoor applications.
	To the best of our knowledge, 
	due to the complexity of anti-collision algorithms,
	comparatively little experimental results of controlling multi-aerial vehicles have been published.
	The efficacy of the proposed method is validated by an indoor experimental scenario.
	Crazyflies, one of the most ubiquitous indoor experimentation platforms, are employed to perform actual experiments.
	Experimental outcomes give us a powerful demonstration of the presented method.
\end{enumerate}
%%
%%

%The rest of this letter is organized as follows.
%%%
%Section \ref{sec:pre} describes algebraic graph theory, a model description of nano-quadrotor systems, formation shape of a group of nano-quadrotor systems, and a collision-free condition.
%%%
%Section \ref{sec:control_algorithm} presents a finite cut-off potential function and a control algorithm.
%%%
%Experimental results are shown in Section \ref{sec:exp_result}.
%%%
%Finally, conclusions are drawn in Section \ref{sec:conclusion}.
%%
%%

%%
{\it Notation: }
The notations $X \geq Y$ and $X > Y$ mean that $X-Y$ is positive semi-definite and positive definite, respectively.
$\otimes$ stands for Kronecker product of two arbitrary-size matrices.
$\Rbb_+$ denotes the set of real positive numbers; 
$\Rbb^n$ and $\Rbb^{n \times m}$ stand for sets of real $n$-dimensional vectors and $n$-row $m$-column matrices, respectively; 
$I_n$ is the $n \times n$ identity matrix;
and $\lambda_{ \min} \left\{ W \right\}$ $\left(\lambda_{\max} \left\{ W \right\}\right)$ represents the minimum (maximum) eigenvalue of matrix $W$.
Next, for a vector $x \in \Rbb^n$,
$\|x\|_2 = \sqrt{x^Tx}$ stands for Euclidean norm in $\Rbb^n$.
%%
%%%%%%%%%%%%%%%%%%%%%%%%%%%%%%%%%%%%%%%%%%%%%%%%%%%%%%%%%%%%%%%%%%%%%%%%%%%%%%%%
\section{Preliminaries}
\label{sec:pre}
%%%%%%%%%%%%%%%%%%%%%%%%%%%%%%%%%%%%%%%%%%%%%%%%%%%%%%%%%%%%%%%%%%%%%%%%%%%%%%%%
\begin{figure}[!t]
	\centering
	\includegraphics[width = \linewidth]{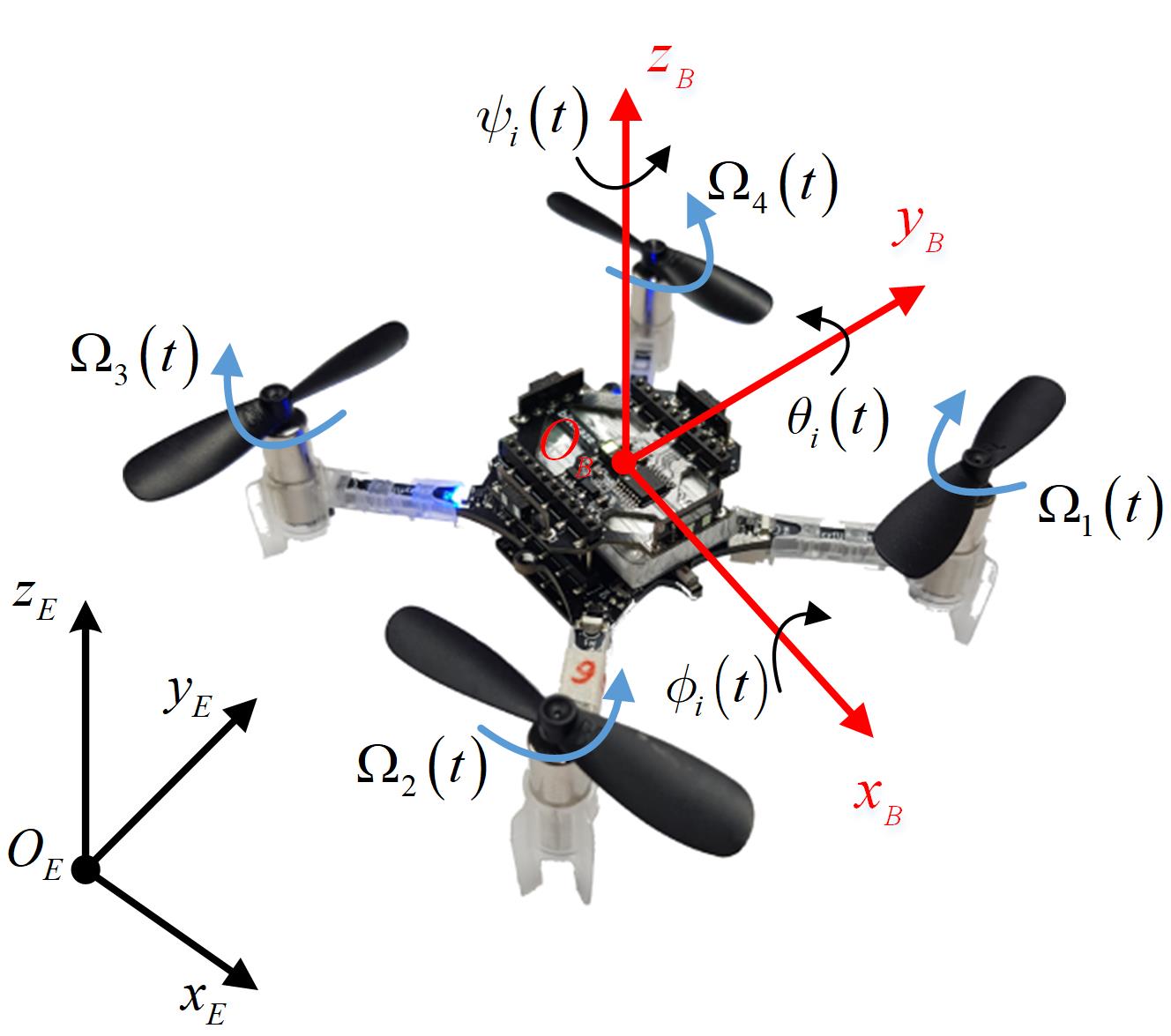}
	\caption{Crazyflie Nano-quadrotor model.}
	\label{Fig:ModelQuad}
\end{figure}

\subsection{Algebraic graph theory}
%------------------------------------------
Let $\Gc = (\Vc, \Ec, \Ac)$ be a weighted digraph with the set of vertices $\Vc = \{1, 2,...,N\}$,
the set of edges $\Ec \subseteq \Vc \times \Vc $, and the weighted adjacency matrix $\Ac  = [a_{ij}]_{{i,j\in \Ec}}$.
For any $(i,j) \in \Ec, ~i\neq j$, the element of the weight adjacency matrix $a_{ij}$ is positive if 
vertices $i$-th and $j$-th can communicate with each other, 
while $a_{ij} = 0$ in the cases of $(i,j) \notin \Ec$ or $i = j$. 
The degree of a vertex $i$-th is denoted as 
$deg_i^{in} = \sum_{j=1}^{n} a_{ij}$, and the degree matrix of the graph $\Gc$ is defined as 
$\Dc = \bs{diag}\big(deg_1^{in}, deg_2^{in},\dots,deg_N^{in}\big)$.
The Laplacian matrix is defined as $\Lc = [\ell_{ij}]_{{i,j\in \Ec}} = \Dc - \Ac$.
Further, $\Gc$ is called an undirected graph if and only if $\Ac$ is a symmetric matrix.
An edge of the undirected graph $\Gc$ is denoted by an unordered pair $(i,j) \in \Ec$. 
The undirected graph is strongly connected if for any pair of vertices, there exists at least a path between two vertices.
The set of all neighbors of the vertex $i$-th is denoted as $\Nc_i = \{j \in \Vc: (i,j) \in \Ec \}$.
%%
%----------------------------
%----------------------------
\begin{figure}[t]
	\centering
	\includegraphics[width= \linewidth]{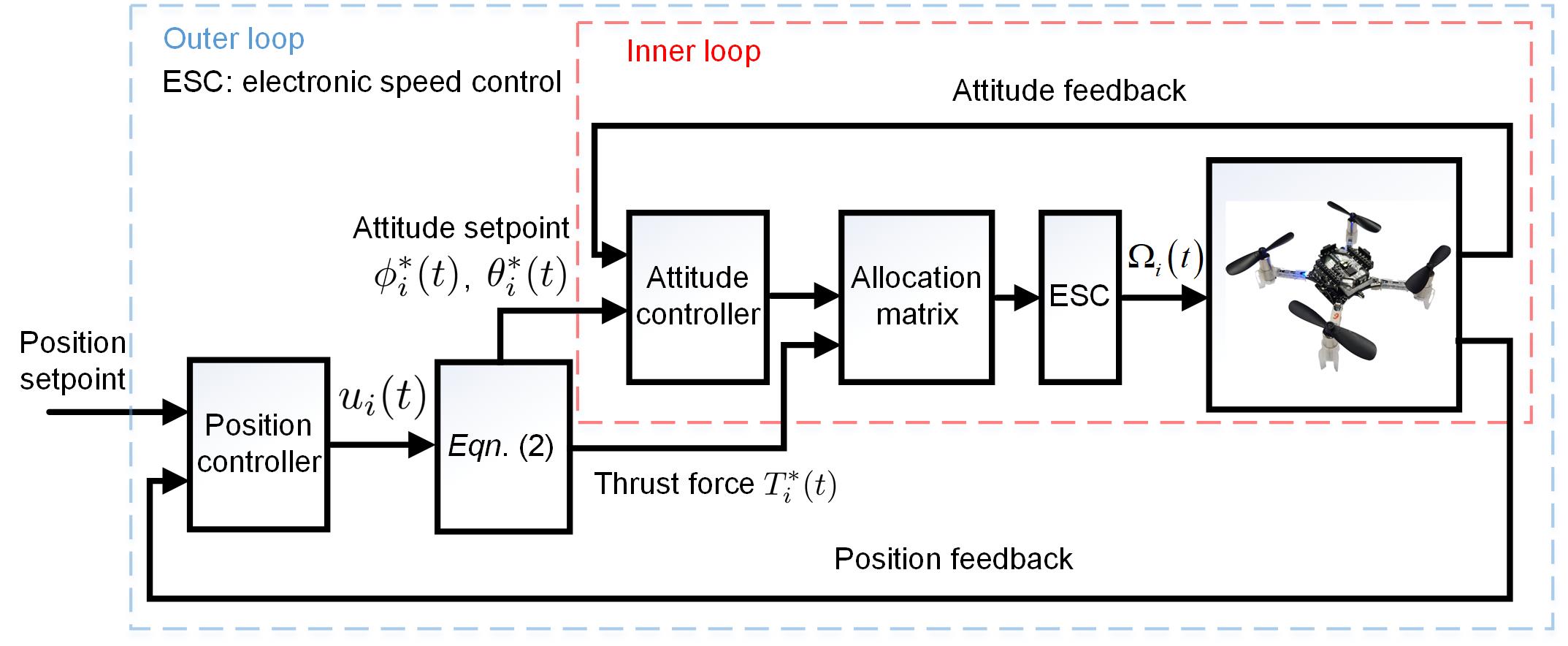}
	\caption{Control diagram of a nano-quadrotor}
	\label{Fig:ctr_diag}
\end{figure}
\subsection{Nano-quadrotor model description}
%----------------------------
%----------------------------
Let us take into account a nano-quadrotor system in Fig. \ref{Fig:ModelQuad} {including:}
body frame $\{Oxyz\}_B$, 
position $p_i(t) = \big[p_{x,i}(t),~p_{y,i}(t),~p_{z,i}(t)\big]^T \in \Rbb^3$ 
, velocity $v_i(t) = \big[v_{x,i}(t),~v_{y,i}(t),~v_{z,i}(t)\big]^T \in \Rbb^3$, and roll/pitch/yaw angles
$\phi_i(t)$/$\th_i(t)$/$\psi_i(t)$ in the Earth-fixed frame $\{Oxyz\}_E$.
{In addition, the rotors $\Omega_1(t)$ and $\Om_3(t)$ ($\Omega_2(t)$ and $\Om_4(t)$) rotates clockwise (anticlockwise) in order to generate a thrust force and moments}.
While the thrust force lifts the vehicle along $z$-axis, the moments rotate the system following $x$-, $y$-, and $z$-axes.
%%
%In this study, only the rotation following $z$-axis is fixed.}
% Em xem cái này có nhất thiết phải ghi vào không vì mình đâu có điều khiển góc đâu nhỉ
%%
Thus, the vehicle is able to move to any positions in the three-dimensional space by adjusting the thrust force and the moments.
{
Inspired by \cite{dong2014time}, this paper successfully} applied the cascade control strategy (see Fig. \ref{Fig:ctr_diag}) that includes inner and outer loops in each controlled quadrotors.
Based on the setups,
the dynamics of the nano-quadrotor in the view of the outer loop can be described as the double integrator:
\begin{align}
\begin{cases}
\dot p_i(t) = v_i(t), \\
\dot v_i(t) = u_i(t), \\ 
\end{cases}
\label{sys:op}
\end{align}
where 
{$u_i(t) = \big[u_{x,i}(t),u_{y,i}(t),u_{z,i}(t)\big]^T \in \Rbb^3$} is a control input of the vehicle, representing the accelerations along $x$-, $y$-, and $z$-axes.
It should be remarked that the nano-quadrotor systems enable us to independently design the outer-loop controller
from which the control input $u_i(t)$ is used to calculated the thrust force $T_i^*(t)$ and the attitude reference $\phi_i^*(t),~\th_i^*(t)$ (refer Eq. \eqref{desiredatt}) toward the inner-loop controller.
Since the fast dynamics of the inner loop,
it can be assumed that the attitude immediately tracks its desired value (see \cite{nguyen2019quadcopter, liao2016distributed, dong2016time} and references therein). Particularly, the relationship among $u_i(t),T_i^*(t),\phi_i^*(t)$ and $\th_i^*(t)$ is given by
\begin{align}
\begin{cases}
T_{i}^*(t) = m_i \sqrt{u_{x,i}^2(t) + u_{y,i}^{2}(t) + \big(u_{z,i}(t) + g\big)^2},  \\
\phi_i^*(t) = \arcsin \frac{m_i u_{x,i}^2\!(t) \sin(\psi_i\!(t)) - m_i u_{y,i}^{2}\!(t) \cos(\psi_i\!(t))}{T_{i}^*(t)},\!\!  \\
\th_i^*(t) = \arctan \frac{u_{x,i}(t)\cos(\psi_i(t)) + u_{y,i}(t)\sin(\psi_i(t))}{u_{z,i}(t) + g}, 
\end{cases}
\label{desiredatt}
\end{align}
where 
$m_i$ and $g$ denote the mass and the gravitational acceleration of the $i$-th nano-quadrotor, respectively.
%%
%------------------------------------
%------------------------------------
\subsection{Formation description}
%------------------------------------
%---------------------
\begin{figure} [t]
	\centering
	\includegraphics[scale=0.8]{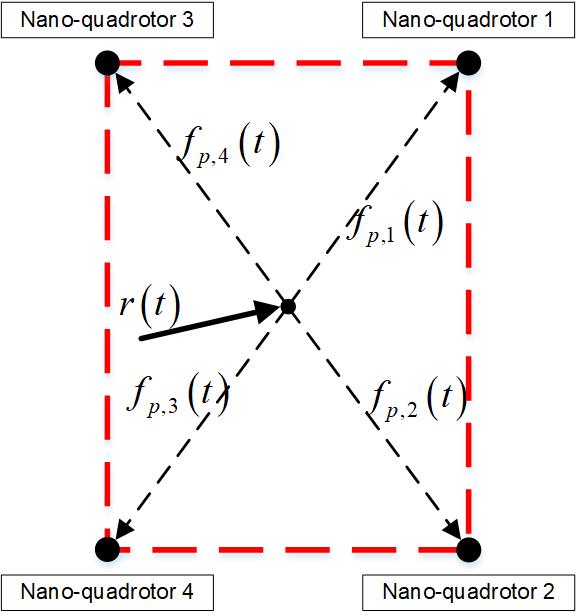}
	\caption{Formation shape of four nano-quadrotors.}
	\label{Fig:FormationExp}
\end{figure}
%----------------------
In this letter, the group of nano-quadrotor systems is considered as the undirected graph $\Gc = (\Vc, \Ec, \Ac)$, in which each nano-quadrotor is a vertex in $\Vc$.
Additionally, a pair $(i,j) \in \Ec$ implies that the $i$-th nano-quadrotor can invoke states of the $j$-th nano-quadrotor and {\it vice versa}.
In an attempt to our control objectives,
the group of $N$ nano-quadrotors is navigated by a proposed control algorithm to reach an anticipated formation. 
Generally, a formation of the multi-quadrotor is considered as a geometric shape in the three-dimensional space that satisfies some prescribed constraints achieved and preserved by the group of nano-quadrotors. 

In what follows, let us define a reference trajectory of the position-based formation $r(t): \Rbb_+ \rightarrow \Rbb^3$ and a formation shape vector of the $i$-th nano-quadrotor $f_{p,i}(t): \Rbb_+ \rightarrow \Rbb^3$, as seen in Fig. \ref{Fig:FormationExp}.
For more details of this formation shape, $f_{p,i}(t)$ is a continuously twice differentiable function, and 
$\|\dot{f}_{p,i}(t)\| = \|{f}_{v,i}(t)\| < \varpi_{fv}$, $\|\ddot{f}_{p,i}(t)\| < \varpi_{fa}$, where $\varpi_{fv}$ and $\varpi_{fa}$ are positive constants. 
Furthermore, 
the $i$-th nano-quadrotor only knows its position in the formation via $f_{p,i}(t)$. 
In the group, at least a nano-quadrotor knows the reference trajectory of the formation $r(t)$.
Let $\delta_i = 1$ if the $i$-th nano-quadrotor knows $r(t)$ and $\delta_i = 0$ otherwise. Let us consider that $r(t)$ satisfies the following
\begin{align}
\dot{r}(t) = v_0 ,
\label{ref}
\end{align}
where $v_0 \in \Rbb^{3}$ is the constant reference velocity of the formation. 
\begin{Definition} \label{def:for}
The multiple nano-quadrotors (\ref{sys:op}) are said to achieve the state formation
specified by the vectors $f_{p,i}(t)$, $i\in \Vc = \{1,2,\dots,N\}$ for any given bounded initial states if
\begin{align}
\lim\limits_{t\rightarrow \infty} \left(p_i(t) - f_{p,i}(t) - r(t) \right) = 0,~i\in \Vc.
\end{align}
\end{Definition}
\begin{Lemma}[\cite{nguyen2020dynamic}] \label{Lem:Mpos}
By letting $\Delta = \bs{diag} \big(\delta_1,~\delta_2,\ldots,\delta_N\big)$ $\in \Rbb^{N\times N}$,
if the graph $\Gc$ is undirected and strongly connect,
the matrices $\Lc + \Delta$ and $\Mc = (\Lc + \Delta)\otimes I_3$ are symmetric positive-definite where $\Lc$ is the Laplace matrix of the graph $\Gc$. 
\end{Lemma}

For the convenience, the tracking errors $e_{p,i}(t),~e_{v,i}(t): \Rbb_+\rightarrow \Rbb^3$ of the $i$-th nano-quadrotor can be defined as follow:
\begin{align}
	e_{p,i}(t) &= p_i(t) - f_{p,i}(t) - r(t),
	\non \\
	e_{v,i}(t) &= v_i(t) - f_{v,i}(t) - v_0.
	\label{sys:tracking}
\end{align}
\begin{Remark}
	When it comes to the complicated trajectories,
	they are generally divided into a sequence of desired points
	that describe the desired position of the formation. 
	Hence,
	the reference trajectory of the group of nano-quadrotors can
	be established by combining many straight lines connected
	two consecutive points in the sequence, i.e., each straight line
	is considered as a constant velocity represented in \eqref{ref}.
\end{Remark}
\begin{Remark}
	Let us consider an obstacle as another agent freely moving in the experimental space. 
	Further, the position of this agent is available in the other agents.
	This letter mainly focus on the scenario in which the obstacle only appears after the group of nano-quadrotors completes the given formation shape.
\end{Remark}

%%%%%%%%%%%%%%%%%%%%%%%%%%%%%%%%%%%%%%%%%%%%%%%%%%%%%%%%%%%%%%%%%%%%%%%%%%%%%%%%
\subsection{Collision-free condition}
\begin{figure}[t]
	\centering
	\includegraphics[scale=0.2]{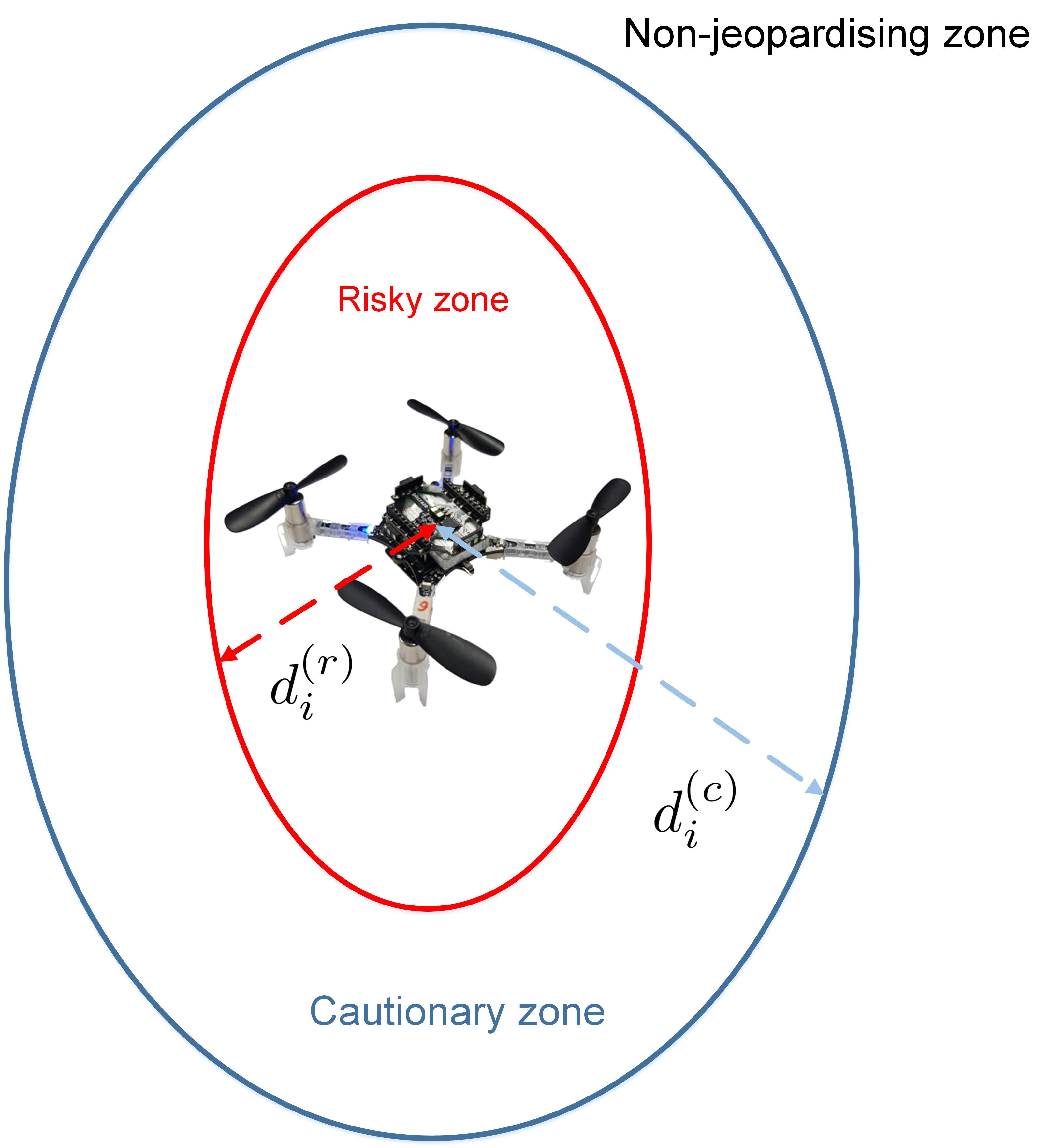}
	\caption{Two virtual zones created surrounding a nano-quadrotor}
	\label{Fig:cf_2zones}
\end{figure}
With the aim of achieving the collision-free, this letter proposes a finite cut-off potential function.
Each nano-quadrotor system possesses its own potential function besides a formation control law. 
While this law manages the anticipated formation flight of the group of multiple nano-quadrotor systems, the control input generated from the proposed potential function guarantees that there are no collisions during the flight time.
How to manipulate the two control inputs can be described by creating three zones, i.e., a risky zone, a cautionary zone, and a non-jeopardising zone in Fig. \ref{Fig:cf_2zones}.
Let us define the radii of the risky and the cautionary zones of the $i$-th nano-quadrotor as $d_{i}^{(r)}$ and $d_{i}^{(c)}$, respectively.
There are three possible circumstances around the $i$-th nano-quadrotor when an object appears in:
\begin{itemize}
\item Non-jeopardising zone: the $i$-th nano-quadrotor solely tracks its desired spot in the anticipated formation shape.
\item Cautionary zone: 
the $i$-th nano-quadrotor prepares for jeopardy the detected object possibly causes.
\item Risky zone: repulsive forces are generated to steer the $i$-th nano-quadrotor away from the jeopardising object.
\end{itemize}
%%
%%%%%%%%%%%%%%%%%%%%%%%%%%%%%%%%%%%%%%%%%%%%%%%%%%%%%%%%%%%%%%%%%%%%%%%%%%%%%%%%
\section{Collision-free formation control}
%%%%%%%%%%%%%%%%%%%%%%%%%%%%%%%%%%%%%%%%%%%%%%%%%%%%%%%%%%%%%%%%%%%%%%%%%%%%%%%%
\label{sec:control_algorithm}
The consideration of collisions to controlling multiple vehicles 
is one of the most challenges for automated driving.
Autonomous systems can be interrupted by some collisions among group members and between vehicles and an obstacle.
The aim of this section is to propose an algorithm to manoeuvre nano-quadrotor systems tracking their given trajectories as well as avoiding collisions.
%%

%%%%%%%%%%%%%%%%%%%%%%%%%%%%%%%%%%%%%%%%%%%%%%%%%%%%%%%%%%%%%%%%%%%%%%%%%%%%%%%%
\subsection{Finite cut-off potential function}
\begin{figure}[!t]
	\centering
	\includegraphics[width=\linewidth]{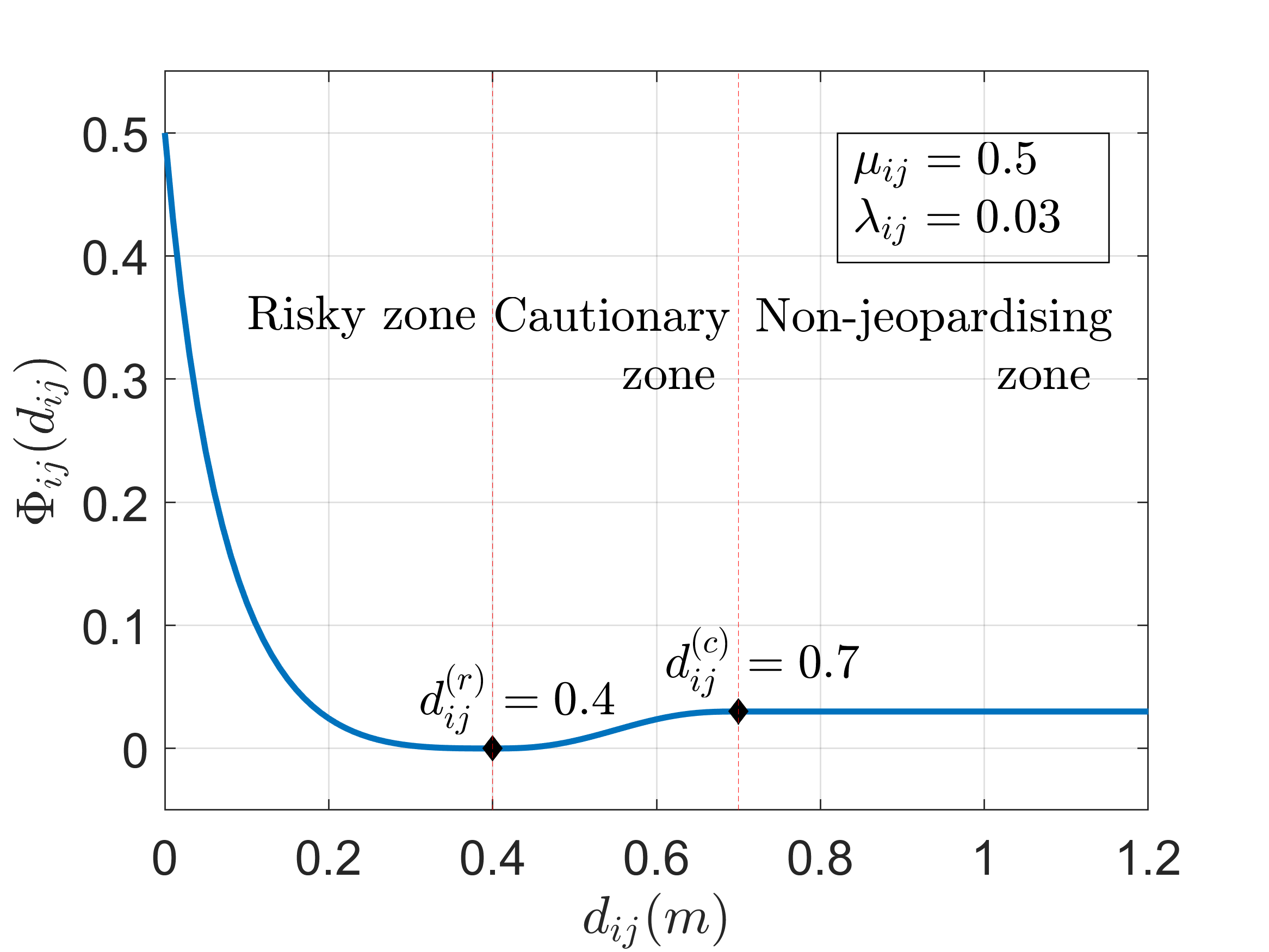}
	\caption{The proposed finite cut-off potential function \eqref{potentialfcn}.}
	\label{Fig:Phi_fcn}
\end{figure}

This part proposes a novel finite cut-off  potential function $\Phi_{ij}(d_{ij}): \Rbb_+ \rightarrow \Rbb_+$ (see Fig. \ref{Fig:Phi_fcn}) that describes the impact of the $j$-th nano-quadrotor on the $i$-th nano-quadrotor ($i \neq j$) as follows:
\begin{align}
\Phi_{ij}(d_{ij}) = f_{ij} (d_{ij}|\mu_{ij}) + \lambda_{ij}  g_{ij} (d_{ij}),
\label{potentialfcn}
\end{align}
where $d_{ij}$ denotes the Euclidean distance between the $i$-th and the $j$-th nano-quadrotors;
two scalars $\lambda_{ij}$ and $\mu_{ij}$ are positive constants such that
\begin{align}
{\lambda_{ij} < \mu_{ij},~ d_{i}^{(r)} < d_{i}^{(c)} < \infty}.
\label{con:lamdamuy}
\end{align}
Next, let $f_{ij} (d_{ij}|\mu_{ij}),~ \forall d_{ij} \in \big[0,~\infty\big)$ as:
\begin{align}
f_{ij} (d_{ij}|\mu_{ij}) = 
\begin{cases}
\displaystyle
\frac{\big(d_{i}^{(r)} - d_{ij}\big)^{3}}{d_{ij} \!+\! d_{i}^{(r)3} \mu_{ij}^{-1}} {,}     & \!\!\text{if}~d_{ij} \!\in\! \big[0,~d_{i}^{(r)}\big], \\
\displaystyle
0{,}   		  & \!\!\text{if}~d_{ij} \!\in\! \left(d_{i}^{(r)},~\infty\right). 
\end{cases}
\label{g_fcn}
\end{align}
\\
\noi
Further, a smooth differentiable step function $g_{ij} (d_{ij}): \Rbb_+ \rightarrow \big[0,~1\big]$ 
is defined \eqref{h_fcn}. 
Based on the definitions of the differentiable step function $g_{ij} (d_{ij})$ and the function $f_{ij} (d_{ij}|\mu_{ij})$, one has
the derivative of the proposed potential function \eqref{dPhi}.
Furthermore, all the properties of the above-defined functions are provided in Appendix A. 
The smoothness and the finiteness of the proposed potential function \eqref{potentialfcn} depicted in Fig. \ref{Fig:Phi_fcn} afford us a great opportunity to implement our method on experimentation platforms. 
\begin{figure*}
	%\rule{\textwidth}{0.2pt}
	\begin{align}
		g_{ij} (d_{ij}) = 
		\begin{cases}
			0{,} 		& \text{if} ~~ d_{ij} \in \big(0, ~d_{i}^{(r)}\big], \\
			\displaystyle \bigg(\frac{d_{ij}-d_{i}^{(r)}}{d_{i}^{(c)}-d_{i}^{(r)}}\bigg)^3  \sum_{k=0}^{2}  \binom{k+2}{k}
			\binom{5}{2-k}\bigg(\frac{d_{i}^{(r)}-d_{ij}}{d_{i}^{(c)}-d_{i}^{(r)}}\bigg)^k{,}	
			& \text{if} ~~ d_{ij} \in \big(d_{i}^{(r)},~ d_{i}^{(c)}\big), \\
			1{,}		& \text{if} ~~ d_{ij} \in \big[d_{i}^{(c)},~ \infty\big).
		\end{cases}
		\label{h_fcn}
	\end{align}
	\begin{align}
		\frac{\partial \Phi_{ij}(d_{ij})}{\partial d_{ij}} =
		\begin{cases}
			\displaystyle
			\frac{-\big(d_{i}^{(r)} - d_{ij}\big)^2 \big(2d_{ij} + 3 d^{(r)3}_{i} \mu_{ij}^{-1} + d_{i}^{(r)}  \big)}{\big(d_{ij} + d^{(r)3}_{i} \mu_{ij}^{-1}\big)^2}{,}    & \text{if}~d_{ij} \in \big[0,~d_{i}^{(r)}\big), \\
			\displaystyle
			\lambda_{ij} \frac{\partial g_{ij} (d_{ij})}{\partial d_{ij}}{,} & \text{if}~d_{ij} \in \big[d_{i}^{(r)},~d_{i}^{(c)}\big), \\
			0{,} 			& \text{if}~d_{ij} \in \big[d_{i}^{(c)},~\infty\big).
		\end{cases}
		\label{dPhi}
	\end{align} 
\end{figure*}
%%%%%%%%%%%%%%%%%%%%%%%%%%%%%%%%%%%%%%%%%%%%%%%%%%%%%%%%%%%%%%%%%%%%%%%%%%%%%%%%
\subsection{Control synthesis}
This section presents a control algorithm for multiple nano-quadrotors with the purpose of steering the group of nano-quadrotors from initial positions to form the anticipated formation shape (see Fig. \ref{Fig:FormationExp}).
Moreover, the multiple nano-quadrotors also follow the given formation trajectory \eqref{ref}, 
and there is no collision between nano-quadrotors in the group and obstacles as well.
Because of such main goals, the designed control algorithm is constructed from two parts, i.e., a collision-free input $u_i^c(t)$ and a formation control input $u_i^f(t)$ as follows:
\begin{align}
	u_i(t) &= u_i^c(t) + u_i^f(t),
	\label{control_input}
	\\
	u_i^c(t) &=  
	\sum_{j \in \Nc_i} %g_{ij} (d_{ij}) 
	\frac{\partial \Phi_{ij}}{\partial d_{ij}}
	\dot d_{ij},
	\non \\
	u_i^f(t) &=
	\Gamma_i
	\Big(-\gamma_p \delta_i e_{p,i}(t)
	- \gamma_v \delta_i e_{v,i}(t)
	\non \\
	&~~~~~~~~
	%~~~~~~~~~~~~~~~~~~~~~~~
	+ \gamma_p \sum_{j \in \Nc_i} \ell_{ij} \big(e_{p,i}(t) - e_{p,j}(t)\big) 
	\non \\
	&~~~~~~~~
	+ \gamma_v \sum_{j \in \Nc_i} \ell_{ij} \big(e_{v,i}(t) - e_{v,j}(t)\big) \Big),\non
\end{align}
\noi
where	$\Gamma_i =  \prod_{j \in \Nc_i} \big(1 - g_{ij} (d_{ij})\big),~\big(0 \leq \Gamma_i \leq 1\big)$,  $\gamma_p$ and $\gamma_v$ are the positive scalars.
\\
{

Let us consider all the circumstances in which there exists an obstacle or another nano-quadrotor inside the detection zone of the $i$-th nano-quadrotor.
First, if the $i$-th nano-quadrotor detects an object (e.g., the $j$-th quadrotor) in its risky zone, i.e., $d_{ij} \leq d_{i}^{(r)}$, the part $u_i^c(t)$ is non-zero to the $i$-th nano-quadrotor.
The control action \eqref{control_input} prefers avoiding collisions with the detected object to forming the formation shape.
In light of A.9 in Appendix A, the purpose of this part is to decrease the value of the function $\Phi_{ij}(d_{ij})$, leading to an increase of the distance from the $i$-th nano-quadrotor to the detected object.
Next, in the second circumstance, the $j$-th nano-quadrotor is detected in the cautionary zone of the $i$-th nano-quadrotor, i.e., $d_{i}^{(r)} < d_{ij} \leq d_{i}^{(c)}$.
Both two terms of the dedicated control input \eqref{control_input} manage the system.
In which, the part $u_i^c(t)$ is capable of keeping the detected object inside the cautionary zone and reducing the repulsive force generated when this object is in the risky zone.
Meanwhile, the part $\Gamma_i$ regulates the impact of the consensus formation control (the part $u_i^f(t)$).
In the last circumstance, there is no object detected in the range of the cautionary zone.
Thanks to the property of the function $\Phi_{ij}(d_{ij})$ (A.9 in Appendix A), only the part $u_i^f(t)$ \eqref{control_input} manoeuvres the $i$-th nano-quadrotor.
The aim of this part is to drive the vehicle such that the group of nano-quadrotors achieves the anticipated formation shape (see Fig. \ref{Fig:FormationExp}).
}
\\
\noi
Next, let us investigate the last circumstance by constructing the closed-loop error dynamic model of the group of multiple nano-quadrotor systems.
In this circumstance, there is no object in the cautionary zone of the $i$-th nano-quadrotor, i.e., $\Gamma_i = 1$.
From the $i$-th nano-quadrotor dynamic model \eqref{sys:op} and the tracking errors \eqref{sys:tracking}, one obtains the following closed-loop error dynamics:
\begin{align}
	e_p(t) =& e_v(t), 
	\non \\
	e_v(t) =& 
	-\gamma_p \delta_i e_{p,i}(t)
	- \gamma_v \delta_i e_{v,i}(t)
	\non \\
	&
	~~~~~~~~~~
	+ \gamma_p 
	\sum_{j \in \Nc_i} \ell_{ij} \big(e_{p,i}(t) - e_{p,j}(t)\big) 
	\non \\
	&
	~~~~~~~~~~
	+ \gamma_v 
	\sum_{j \in \Nc_i} \ell_{ij} \big(e_{v,i}(t) - e_{v,j}(t)\big) .
	\label{sys:closedloop}
\end{align}
Then, the closed-loop error dynamic of $N$ nano-quadrotors follows:
\begin{align}
	\bs{\dot e}_p(t) &= \bs{\dot e}_v(t), 
	\non	\\
	\bs{\dot e}_v(t) &= 
	\gamma_p \Mc \bs{e}_p(t) 
	- \gamma_v \Mc \bs{e}_v(t),
	\label{sys:N_closeloop}
\end{align}
where
$\bs{e}_p(t) = \big[e_{p,1}^T(t),~e_{p,2}^T(t),\ldots,e_{p,N}^T(t)\big]^T$ and
$\bs{e}_v(t) =$ $\big[e_{v,1}^T(t),~e_{v,2}^T(t),\ldots,e_{v,N}^T(t)\big]^T$.
\\
{
The following theorem provides a formation control algorithm for the multiple nano-quadrotor systems, by which there are no collisions among the group members and obstacles}. 
\begin{Theorem} \label{theo:theo1}
	{\it (Proof in Appendix A)}
	Let us consider the multiple nano-quadrotor systems \eqref{sys:op}. Suppose that the graph $\Gc$ is undirected and strongly connected, and there exist positive scalar coefficients $\lambda_{ij}, \mu_{ij}, \gamma$, $\gamma_p$, $\gamma_v$, $\th_p$, and $\th_v$ such that:
	\begin{align}
		& \mu_{ij} > \Phi_{ij}(d_{ij}(0)),~ \forall i \in \Vc,~j \in \Nc_i,
		\non \\
		&0 < \gamma_p - \th_p,
		%\label{thm1:con1}
		\non \\
		&0 < \left(\gamma_v - \th_v\right) \Mc^2 - \gamma \Mc,
		%\label{thm1:con2}  
		\non \\
		&0 < \left(\gamma_p + \gamma_v \gamma\right) \Mc^2 - \gamma^2 \Mc,
		%\label{thm1:con3}
		\label{theo:con}
	\end{align}
	where $\Mc$ is defined in Lemma 1.
	Then, under the control input \eqref{control_input}, the following statements hold:
	\begin{enumerate}[i)]
		\item The group of nano-quadrotors is collision-free, and
		\item  State formation in Definition \ref{def:for} is achieved. 
	\end{enumerate}
\end{Theorem}
%%
%%%%%%%%%%%%%%%%%%%%%%%%%%%%%%%%%%%%%%%%%%%%%%%%%%%%%%%%%%%%%%%%%%%%%%%%%%%%%%%%
\section{Experimental Results}
\begin{figure*} [!t]
	\centering
	\includegraphics[width = \textwidth]{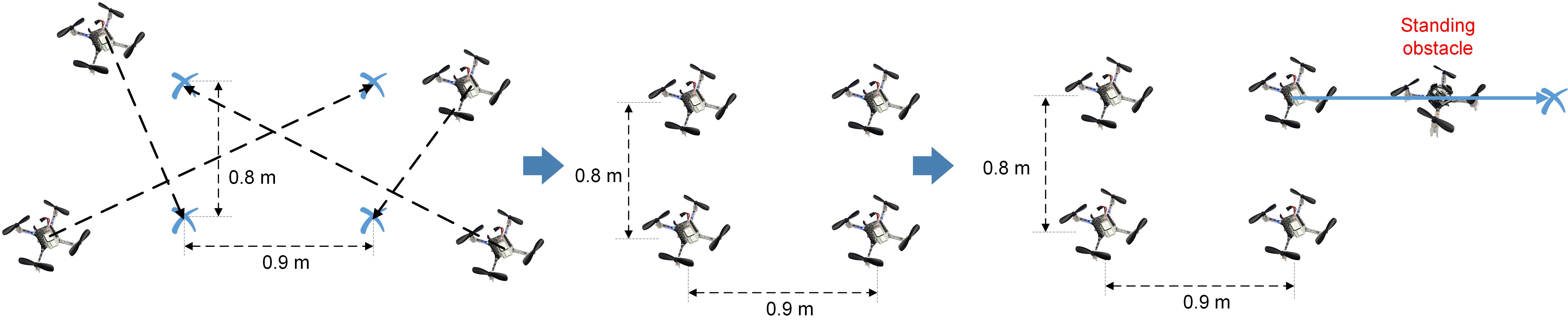}
	\caption{Experimental scenario of five Crazyflies.}
	\label{Fig:scenario_setup}
	\begin{subfigure}{0.46\textwidth}
		\includegraphics[width = \textwidth]{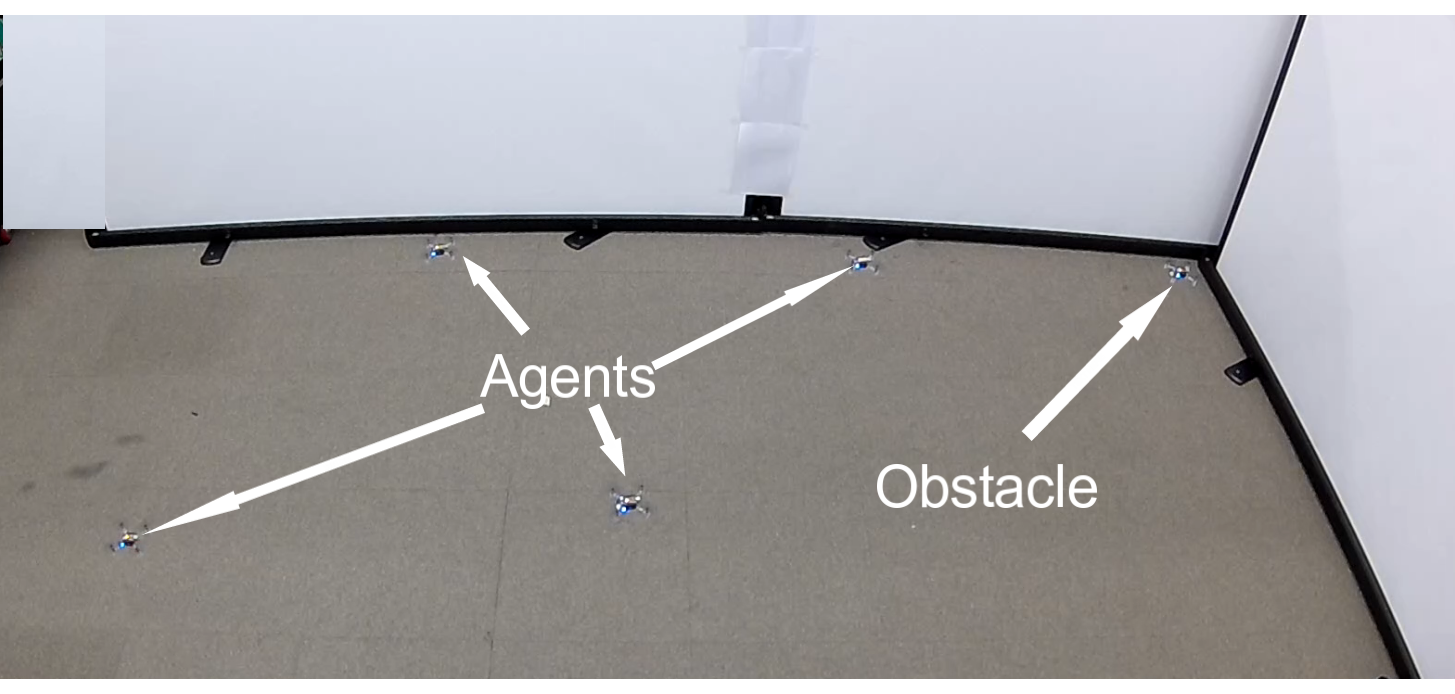}
		\caption{}
		\label{Fig:test_a}
	\end{subfigure}
	\begin{subfigure}{0.46\textwidth}
		\includegraphics[width = \textwidth]{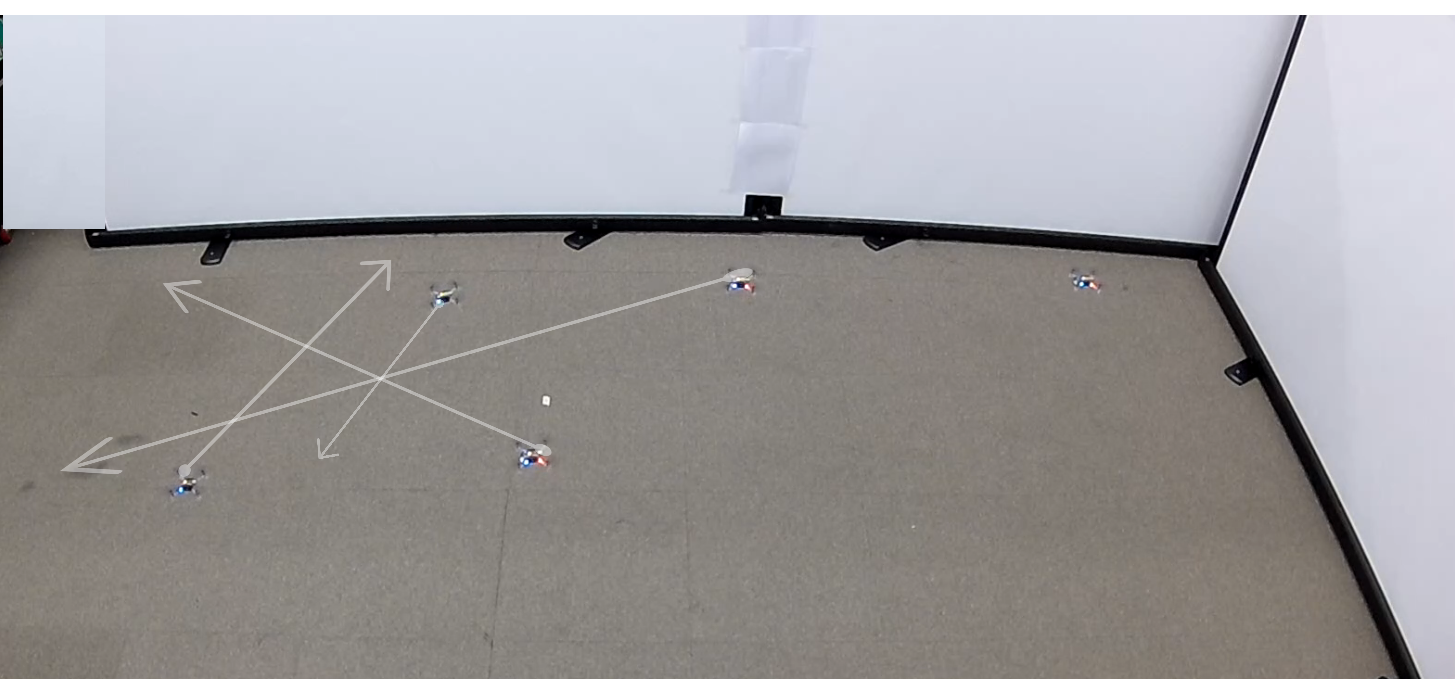}
		\caption{}
		\label{Fig:test_b}
	\end{subfigure}
	\begin{subfigure}{0.46\textwidth}
		\includegraphics[width = \textwidth]{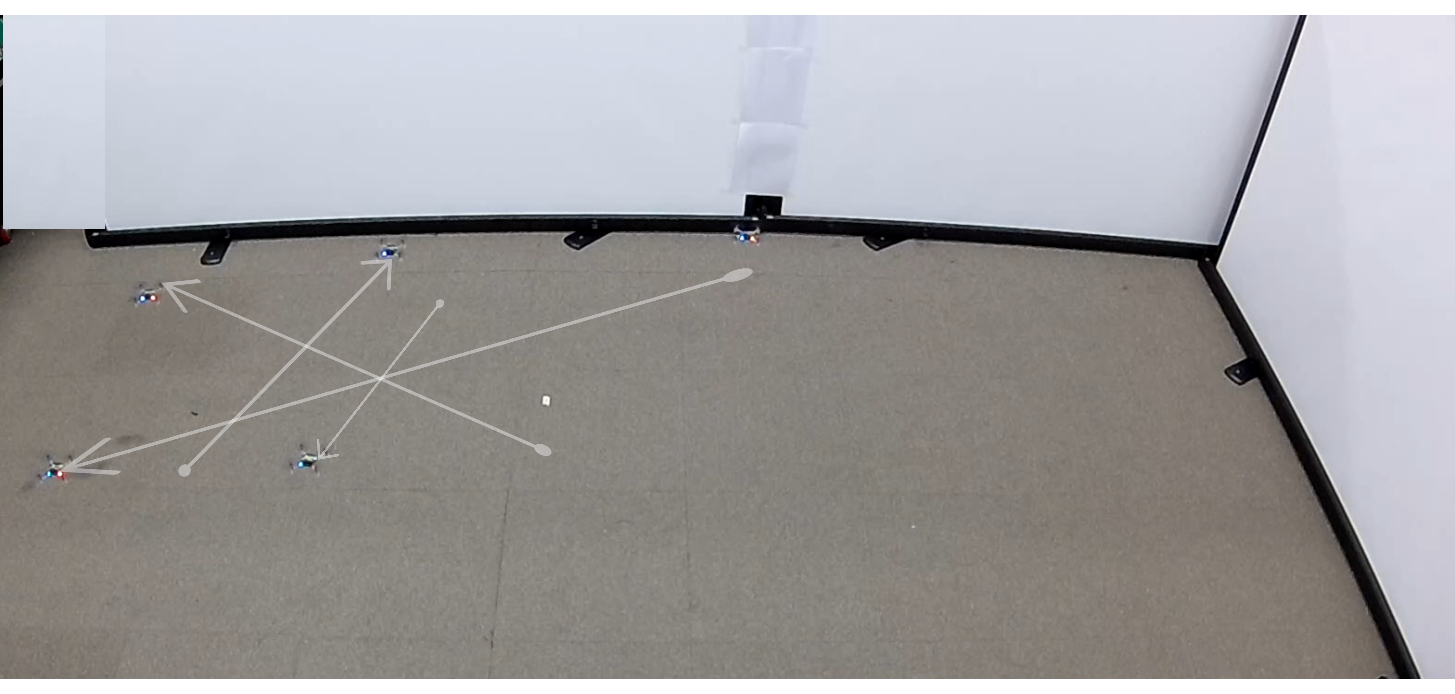}
		\caption{}
		\label{Fig:test_c}
	\end{subfigure}
	%%
%	\hspace{0.4cm}
	%%
	\begin{subfigure}{0.46\textwidth}
		\includegraphics[width = \textwidth]{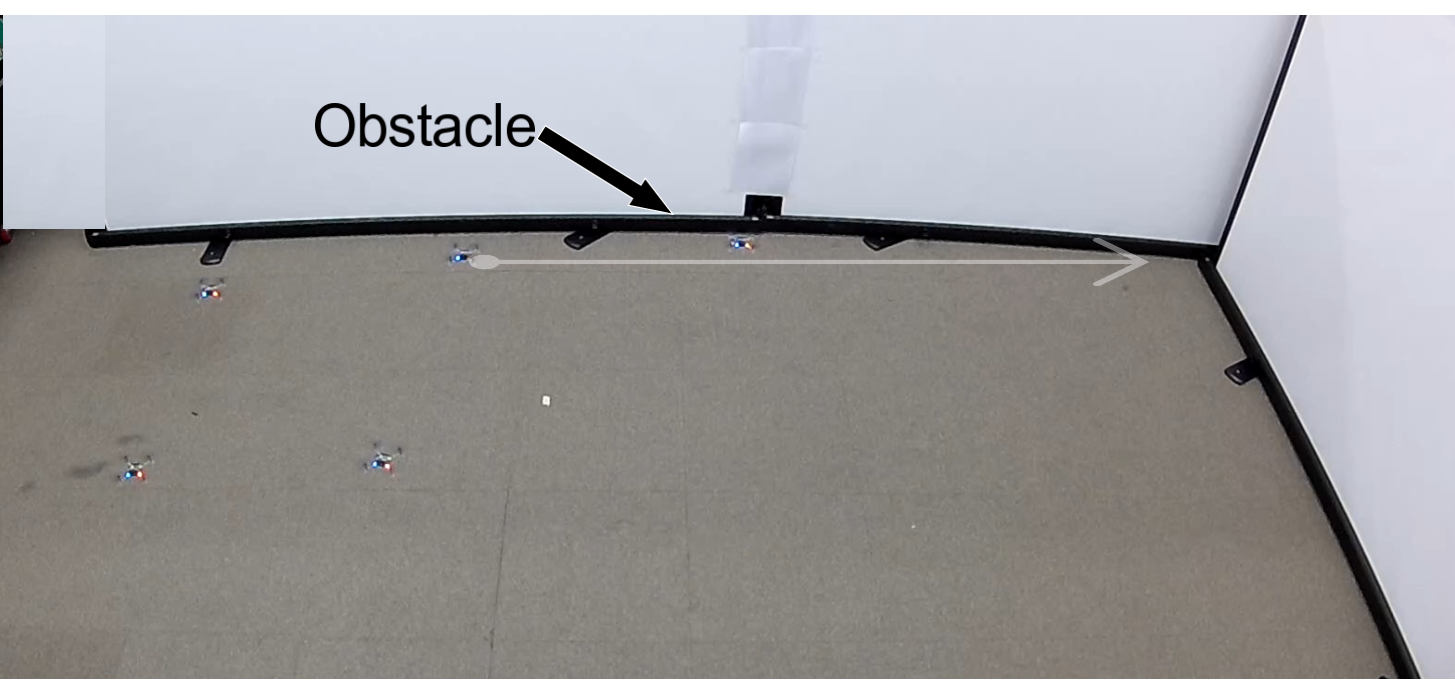}
		\caption{}
		\label{Fig:test_d}
	\end{subfigure}
	\caption{Actual tests of five Crazyflies: (a) taking off at ground locations, (b) moving to desired spots of an anticipated formation shape, (c) achieving the formation shape, (d) tracking a straight trajectory.}
\end{figure*}
\label{sec:exp_result}
\begin{figure}[htpb]
	\centering
	\includegraphics[width=\linewidth]{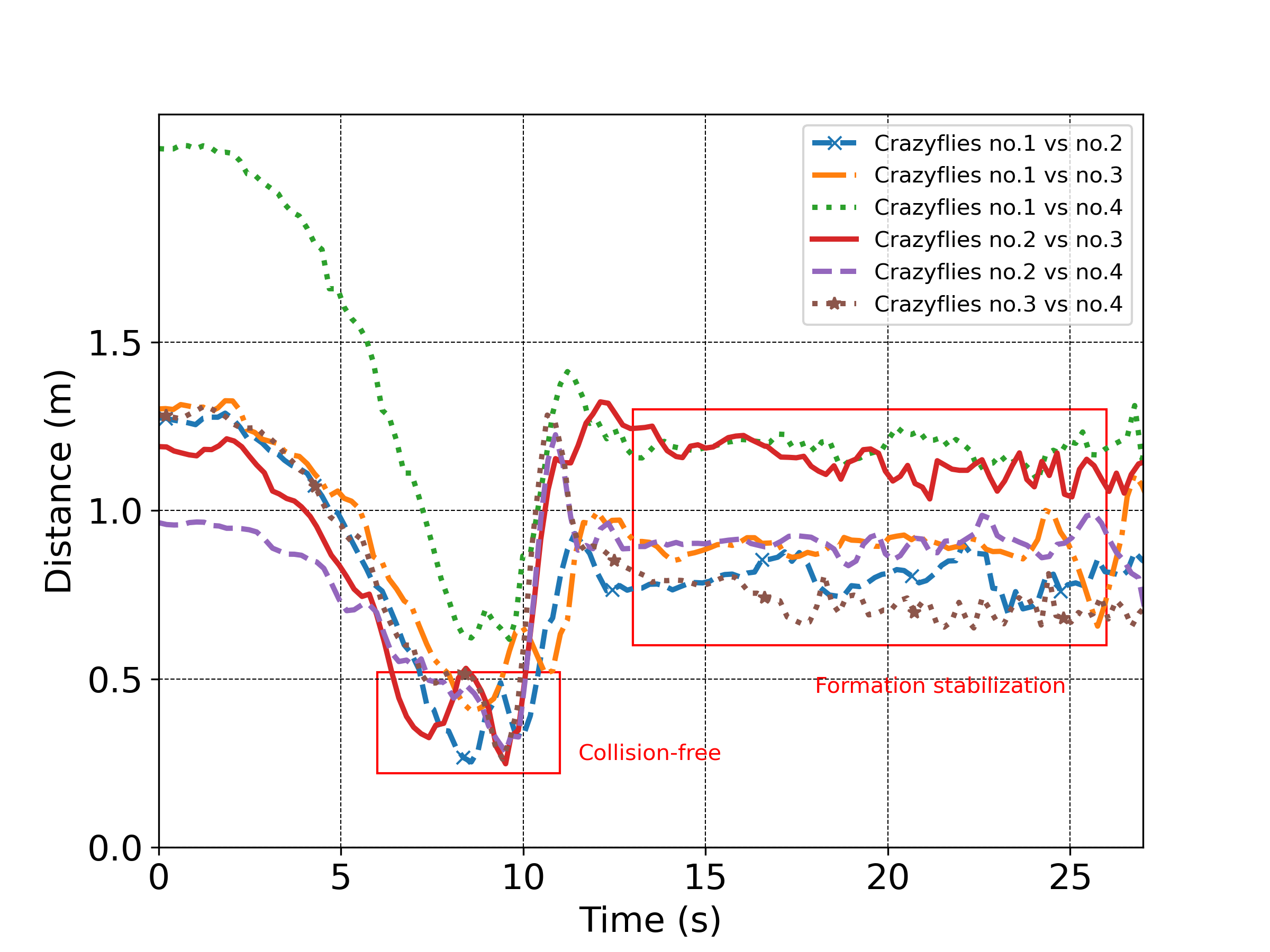}
	\caption{Distances between couples of agent-Crazyflies.}
	\label{Fig:dis_couple}
\end{figure}
\begin{figure}[htpb]
	\centering
	\includegraphics[width=\linewidth]{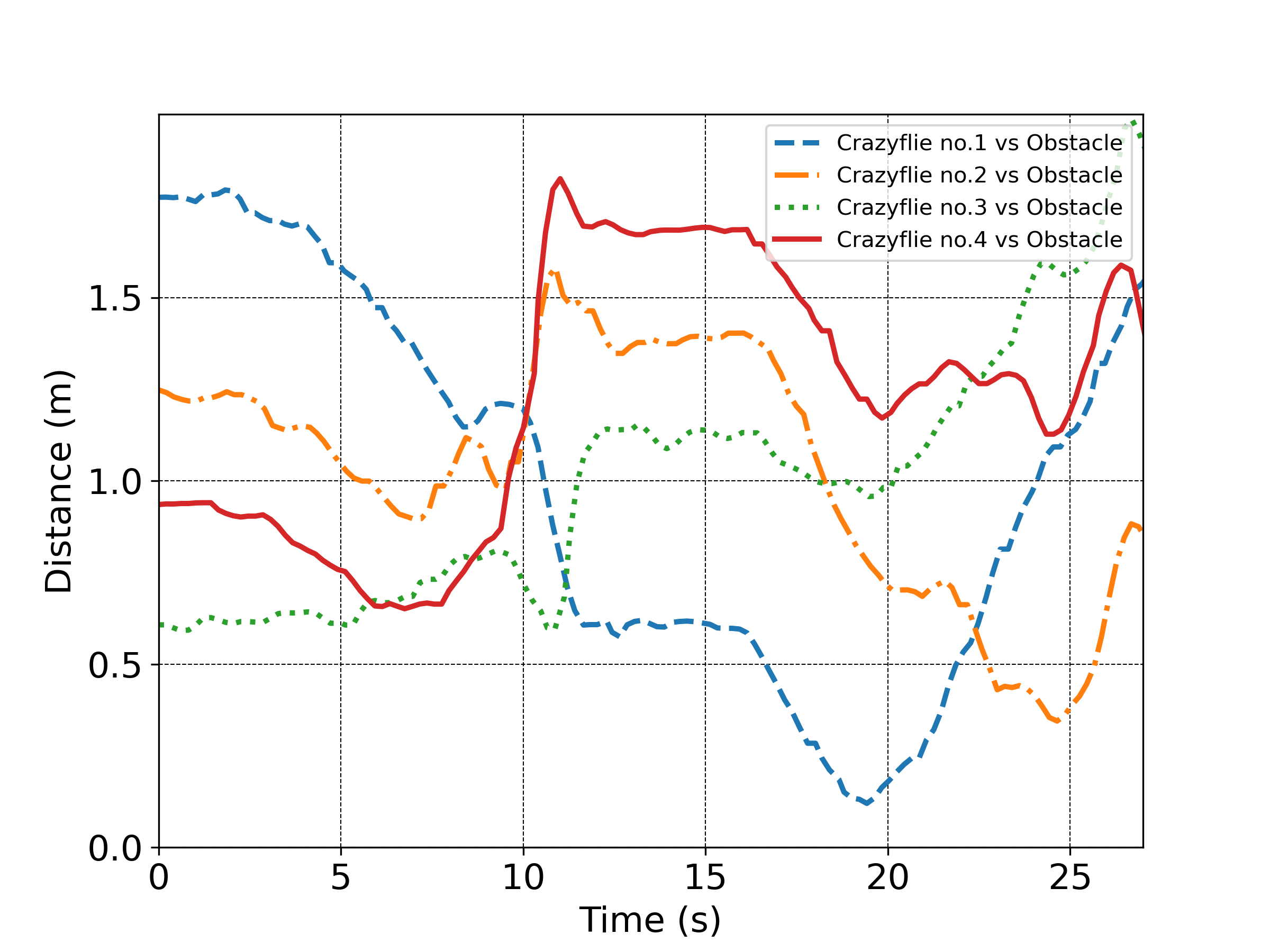}
	\caption{Distances between an agent-Crazyflies and an obstacle-Crazyflies.}
	\label{Fig:dis_ob}
\end{figure}
\begin{figure}[htpb]
	\centering
	\includegraphics[width=\linewidth]{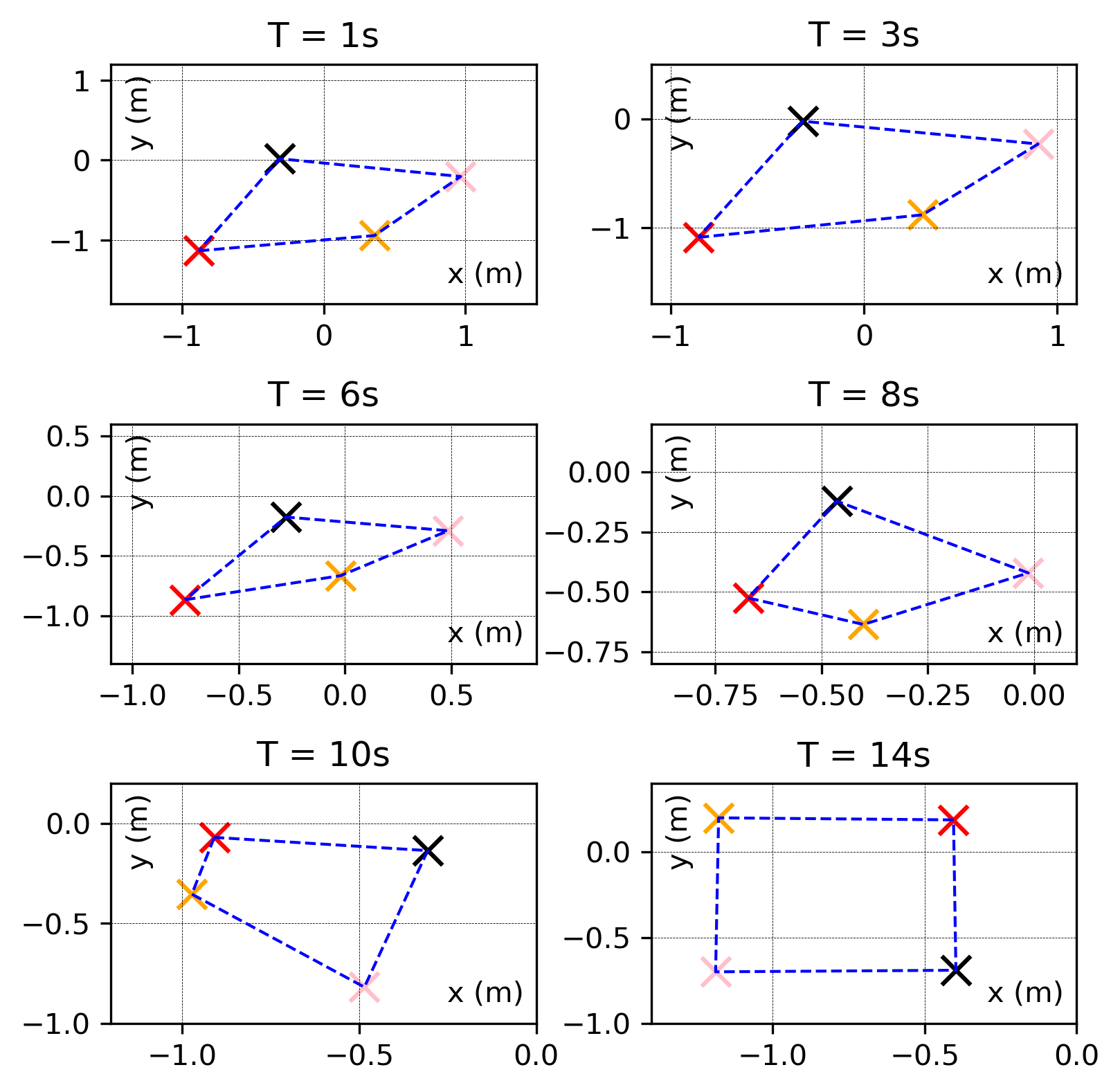}
	\caption{Position snapshots of agent-Crazyflies.\\}
	\label{Fig:snapshot}
\end{figure}
\begin{figure}[htpb]
	\centering
	\includegraphics[width=\linewidth]{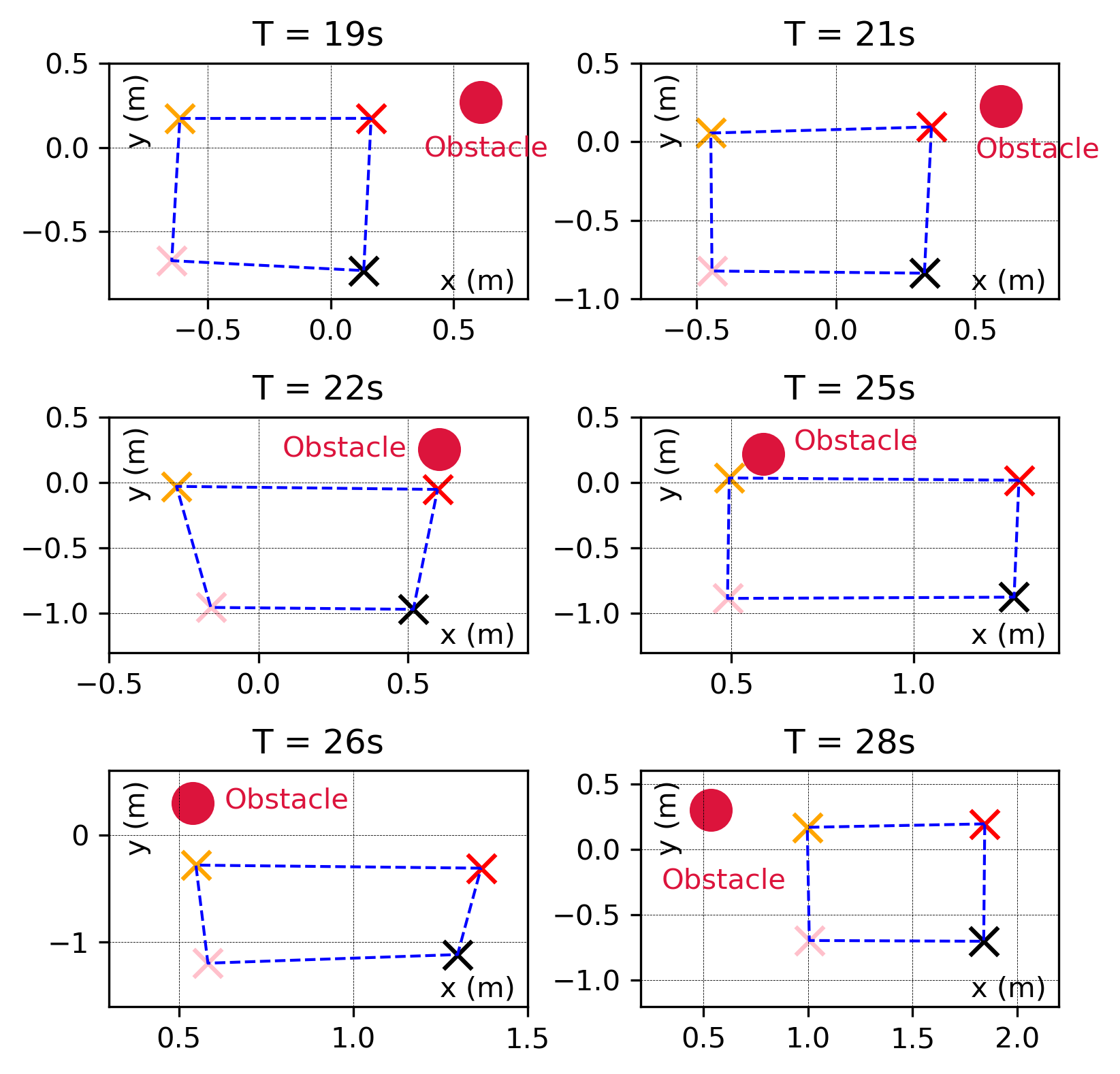}
	\caption{Position snapshots of agent-Crazyflies moving through an obstacle.}
	\label{Fig:snapshot1}
\end{figure}
By carrying out flight experiments, this section shows a powerful demonstration of the proposed algorithm introduced in Section \ref{sec:control_algorithm}.
The multiple nano-quadrotors utilized in our flight experiments were Crazyflies (see Fig. \ref{Fig:ModelQuad}) developed by Bitcraze \cite{cf212020}.
To be specific,
Four Crazyflies were employed as agents to form a given rectangular shape in Fig. \ref{Fig:FormationExp} and another Crayzlife was used as an obstacle.
In addition, each Crazyflie was equipped with an indoor positioning receiver to perceive its position in an experimental room.
This system has afforded us a Python library to send outer-loop control commands via a wireless communication network.

In order to validate our method, the scenario was constructed by three main stages in Fig. \ref{Fig:scenario_setup} (total of 28-second flight time).
At the first stage, the four agent-Crazyflies were placed at initial positions (see Fig. \ref{Fig:test_a}).
Trajectories of the four agent-Crazyflies from such initial positions to their desired spots were designed such that they crossed each other (see Fig. \ref{Fig:test_b}).
This setup possibly resulted in collision points and offered us a golden opportunity to verify the proposed algorithm.
The aim of the four agent-Crazyflies at the next stage was to form the anticipated formation shape in Fig. \ref{Fig:FormationExp} without any collisions among the group members (see Fig. \ref{Fig:test_c}).
At the final stage, the four agent-Crazyflies tracked a straight trajectory in Fig. \ref{Fig:test_d}.
There was another Crazyflie described as an obstacle at a middle point of the trajectory.
For more details, the initial positions of the four agent-Crazyflies and the obstacle-Crazyflie were placed at different locations in the experimental room, while their desired formation shapes of all the agent-Crazyflies were selected as: 
	$f_{p,1} = [0.4,0.45,~0.0]^T$ , $p_{p,2} = [-0.4,~0.45,~0.0]^T$, $p_{p,3} = [0.4,~-0.45,~0.0]^T$, and $p_{p,4} = [-0.4,~-0.45,~0.0]^T$, 	$p_{ob} = [0.2,~0.2,~0.4]^T$.
Based on Theorem \ref{theo:theo1}, all the coefficients were chosen as: $\gamma = 0.2, \gamma_v = 3, \gamma_p = 2, \th_p = 1.3, \th_v = 1.3, \mu_{ij} = 0.5, \lambda_{ij} = 10^{-3}, d_i^{(r)} = 0.4$m, and $d_i^{(c)} = 0.7$m.
Laplacian matrix was selected such that each nano-quadrotor is able to communicate with the other nano-quadrotors.
The reference velocity was set $v_0 = 0.4$ m/s.

As mentioned above, 
%% Nên nói tổng số thời gian mô phỏng trước, có thể đưa lên đoạn graph trên cũng đc nhé
%while the third one operated for the rest of the experiment time until a landing.
%%
Figs. \ref{Fig:snapshot} and \ref{Fig:snapshot1} illustrated position snapshots of the four agent-Crazyflies in the 28-second flight time.
The four agent-Crazyflies smoothly moved to their desired spots until the $6^{\text{th}}$ second when they sensed near objects that might occur collisions.
In the following six seconds, collision avoidance was witnessed in Fig. \ref{Fig:snapshot}.
Distances between couples of the agent-Crazyflies were shown in Fig. \ref{Fig:dis_couple}.
When the agent-Crazyflies moved near the collision points, these distances gradually decreased and surpassed $d_i^{(r)} = 0.4$m at the $6^{\text{th}}$ second.
Thanks to the advances of the proposed potential function, the part $u^c(t)$ of the control input \eqref{control_input} was fairly activated to create repulsive forces among the agent-Crazyflies.
The activated repulsive forces smoothly pushed the agent-Crazyflies far away from the collision points.
These forces guaranteed a safe distance among the agent-Crazyflies.
Further, the lines in Fig. \ref{Fig:dis_couple} from the $6^{\text{th}}$ to the $14^{\text{th}}$ seconds never hit the zero-line, clearly illustrating no collisions among the agent-Crazyflies.
After the 14-second fight time, all the agent-Crazyflies reached their desired spots in the formation shape (see Fig. \ref{Fig:snapshot}).
The consensus formation flight was achieved.
At the next stage, the formation trajectory was built as a straight line in Fig. \ref{Fig:test_d}.
Due to the presence of the obstacle, the four agent-Crazyflies changed their movements to avoid this obstacle-Crazyflie.
This execution was shown from the $19^{\text{th}}$ to the $28^{\text{th}}$ seconds in Fig. \ref{Fig:snapshot1}.
Fig. \ref{Fig:dis_ob} illustrated that distances between the four agent-Crazyflies and the obstacle-Crazyflie never hit the zero-line.
This highly confirmed that no obstacle collision occurs.
{Please refer to a video at
\href{https://bit.ly/37nno4t}{{\bf https://bit.ly/37nno4t}}
for further observations of the actual experiments.}

%%%%%%%%%%%%%%%%%%%%%%%%%%%%%%%%%%%%%%%%%%%%%%%%%%%%%%%%%%%%%%%%%%%%%%%%%%%%%%%%
\section{Conclusion}
\label{sec:conclusion}
This letter has addressed the formation control of multiple nano-quadrotor systems regarding collisions among group members and between vehicles and an obstacle.
The collision-free formation protocol has been presented to deal with such a problem.
The effectiveness and validity of the proposed method have been vividly demonstrated by the actual experiments via Crazyflies nano-quadrotors.
In future work, we plan to study the control of multiple nano-quadrotor systems under cyber-attacks in which some agents lose control and are managed by attackers.

%\section*{ACKNOWLEDGMENT}
%%%%%%%%%%%%%%%%%%%%%%%%%%%%%%%%%%%%%%%%%%%%%%%%%%%%%%%%%%%%%%%%%%%%%%%%%%%%%%%%

\section*{APPENDIX A \\ Properties of the proposed potential functions}
%%%%%%%%%%%%%%%%%%%%%%%%%%%%%%%%%%%%%%%%%%%%%%%%%%%%%%%%%%%%%%%%%%%%%%%%%%%%%%%%
\!\!\!\!\!\!$\bullet$ Properties of the function $f_{ij} (d_{ij}|\mu_{ij})$
\begin{itemize}
	\item[A.1\!] $0 \leq f_{ij} (d_{ij}|\mu_{ij}) \leq \mu_{ij}$
	\\
	\item[A.2\!] Derivative $\displaystyle \frac{\partial f_{ij} (d_{ij}|\mu_{ij})}{\partial d_{ij}}$
	exists and is continuous.
\end{itemize}
$\bullet$  Properties of the function $g_{ij} (d_{ij})$
\begin{itemize}
\item[A.3\!] $g_{ij} (d_{ij})$ is continuous and differentiable  
$\forall d_{ij} \in \big(0,~\infty \big)$,
$0 < g_{ij} (d_{ij}) < 1,~ \forall d_{ij} \in \big(d_{i}^{(r)},~ d_{i}^{(c)}\big)$, \label{hprop1}
\\
\item[A.4\!] $ \frac{\partial g_{ij} (d_{ij})}{\partial d_{ij}}$ is continuous $\forall d_{ij} \in \big(0,~\infty\big)$,
\label{hprop3}
\\
\item[A.5] $ \frac{\partial g_{ij} (d_{ij})}{\partial d_{ij}} >0,~ \forall d_{ij} \in \big(d_{i}^{(r)},~ d_{i}^{(c)}\big) $, $ \frac{\partial^k g_{ij} (d_{ij})}{\partial d_{ij}^k} = 0$,  $\forall d_{ij} \in \big(0,~d_{i}^{(r)}\big] \cup \big[d_{i}^{(c)},~\infty\big)$. \label{hprop5}
\end{itemize}

$\bullet$ Properties of the function $\Phi_{ij}(d_{ij})$
\begin{itemize}
	\item[A.6\!] $ 0 \leq \Phi_{ij}(d_{ij}) < \mu_{ij},~\forall d_{ij} \in \big(0,~\infty\big)$,
	\\
	\item[A.7\!] $ \max_{d_{ij} \geq 0} \Phi_{ij}(d_{ij}) = \Phi_{ij}(0) = \mu_{ij}$,
	\\
	\item[A.8\!] $ \frac{\partial \Phi_{ij}(d_{ij})}{\partial d_{ij}}$ is bounded and continuous \\
	$\forall d_{ij} \in \big(0,~\infty\big)$,
	\\
	\item[A.9\!] $ \frac{\partial \Phi_{ij}(d_{ij})}{\partial d_{ij}} < 0,\forall d_{ij} \in \big(0,d_{i}^{(r)}\big)$,	\label{Pro:phi_4}
	$ \frac{\partial \Phi_{ij}(d_{ij})}{\partial d_{ij}} > 0,~\forall d_{ij} \in \big(d_{i}^{(r)},~d_{i}^{(c)}\big)$, 
	$ \frac{\partial \Phi_{ij}(d_{ij})}{\partial d_{ij}} = 0,~\forall d_{ij} \in \big[d_{i}^{(c)},~\infty\big)$.
\end{itemize}

\section*{APPENDIX B \\ Proof of Theorem 1}
%%%%%%%%%%%%%%%%%%%%%%%%%%%%%%%%%%%%%%%%%%%%%%%%%%%%%%%%%%%%%%%%%%%%%%%%%%%%%%%%
In the scope of this study, collision points are mainly addressed when nano-quadrotors move from their initial positions to their desired spots in the formation shape (Fig. \ref{Fig:FormationExp}).
Collisions between vehicles and an obstacle are also considered after the group members reach their anticipated formation.	
The part $u_i^c(t)$ of the proposed control law \eqref{control_input} guarantees that nano-quadrotors prefer avoiding jeopardizing objects inside their cautionary zones to forming the given formation shape.
This means that a nano-quadrotor tends to automatically change its direction with a view to placing all the detected objects outside its cautionary zone, which achieves i).
\\
By utilizing Schur's complement, the condition \eqref{theo:con} and Lemma \ref{Lem:Mpos} hold that:
	\begin{align}
	\Pbb = 
	\ba{cc}
	(\gamma_p + \gamma_v \gamma)\Mc^2 & \gamma \Mc\\
	%\hdashline
	\gamma \Mc &\Mc
	\ea > 0. \label{blockmatrixpos}
	\end{align}
	Next, let us take a Lyapunov function candidate as follows:
	\begin{align}
	V(t) = \big[\bs{e}_p^T(t),~\bs{e}_v^T(t)\big]
	\Pbb \big[\bs{e}_p^T(t),~\bs{e}_v^T(t)\big]^T
	\label{theo:V}
	\end{align}
	The time-derivative of \eqref{theo:V} along with the solution of \eqref{sys:N_closeloop} is represented by:
	\begin{align}
	\dot V(t) = 
	&(\gamma_p + \gamma_v \gamma)\bs{e}_p^T(t) \Mc^2 \bs{e}_v(t) 
	+ \gamma \bs{e}_v^T(t) \Mc \bs{e}_v(t) 
	\non \\ 
	&+ \gamma \bs{e}_p^T(t) \Mc \bs{\dot e}_v(t) 
	+ \bs{e}_v^T(t)  \Mc \bs{\dot e}_v(t)
	\non \\
	=
	&\!-\! \gamma_p \gamma \bs{e}_p^T \Mc^2 \bs{e}_p(t) \!-\!  \bs{e}_v^T(t) (\gamma_v \Mc^2 \!-\! \gamma \Mc) \bs{e}_v(t) 
	\non \\
	=
	& - \big[\bs{e}^T_p(t),~\bs{e}_v^T(t)\big]
	\Mbb \big[\bs{e}_p^T(t),~\bs{e}_v^T(t)\big]^T,
	\label{theo:Vdot}
	\end{align}
	where 
	\begin{align}
	\Mbb = \ba{cc}
	\!\!\!\gamma (\gamma_p \!-\! \th_p)\Mc^2 &0\\
	%\hdashline
	0              &(\gamma_v \!-\! \th_v) \Mc^2 \!-\! \gamma \Mc
	\ea. \non
	\end{align}	
	By invoking the conditions \eqref{theo:con}, the matrix $\Mbb$ is positive definite. From the fact that $\Mbb \geq \frac{\lambda_{min}(\Mbb)}{\lambda_{max}(\Pbb)}$, where $\Pbb$ is also a positive finite matrix, let us choose:
	\begin{align}
	\zeta = \frac{2\lambda_{min}(\Mbb)}{\lambda_{max}}. \non
	\end{align}
	Then, \eqref{theo:Vdot} gives:
	\begin{align}
	\dot V(t) \leq - \zeta V(t),
	\label{theo:Vdot2}
	\end{align}
	for $\forall t \geq 0$. Therefore, in light of the comparison lemma \cite[pp. 102]{khalil2002nonlinear}, one has:
	\begin{align}
	0 \leq V(t) \leq V(0) e^{-\zeta t}. \non
	\end{align}
	Thanks to the assumptions of a given reference trajectory, initial tracking errors are bounded, leading to $V(0) < \infty$. 
	For all bounded $V(0)$, $\displaystyle \lim_{t \rightarrow \infty} V(0) e^{-\zeta t}= 0$. 
	This derives that $\displaystyle \lim_{t \rightarrow \infty} V(t) = 0$.
	Because of the positive definite matrix $\Pbb$ \eqref{blockmatrixpos}, 
	$\displaystyle \lim_{t \rightarrow \infty} \bs{e}_p(t) = 0$ and 
	$\displaystyle \lim_{t \rightarrow \infty} \bs{e}_v(t) = 0$, which proves ii).

%------------------------------------------
%------------------------------------------
\bibliographystyle{IEEEtran}
\bibliography{IEEEabrv,mybibfile}
%------------------------------------------

\end{document}